\documentclass[letter]{aa}  

\usepackage{graphicx}
\usepackage{txfonts}
\usepackage{pdfpages}
\usepackage{tabularx}
\usepackage{amssymb,amsmath}
\usepackage{longtable}
\begin{document}

   \title{The ALMA-PILS survey: First detection of nitrous acid (HONO) in the interstellar medium}

   \titlerunning{First detection of HONO in the interstellar medium}
   
   \author{A. Coutens\inst{1} \and
   N. F. W. Ligterink\inst{2} \and 
   J.-C. Loison\inst{3} \and
   V. Wakelam\inst{1} \and 
   H. Calcutt\inst{4} \and
   M. N. Drozdovskaya\inst{2} \and
   J. K. J\o rgensen\inst{5} \and \\
   H. S. P. M\"{u}ller\inst{6} \and 
   E. F. van Dishoeck\inst{7,8} \and
   S. F. Wampfler\inst{2}
              }

   \institute{Laboratoire d'astrophysique de Bordeaux, Univ. Bordeaux, CNRS, B18N, all\'ee Geoffroy Saint-Hilaire, 33615 Pessac, France\\
              \email{audrey.coutens@u-bordeaux.fr}
         \and
             Center for Space and Habitability (CSH), University of Bern, Sidlerstrasse 5, 3012 Bern, Switzerland
         \and
             Institut des Sciences Mol\'eculaires (ISM), CNRS, Universit\'e Bordeaux, 351 cours de la Lib\'eration, F-33400, Talence, France
         \and
         Department of Space, Earth and Environment, Chalmers University of Technology, 41296 Gothenburg, Sweden
         \and 
         Centre for Star and Planet Formation, Niels Bohr Institute and Natural History Museum of Denmark, University of Copenhagen,
\O ster Voldgade 5-7, 1350 Copenhagen K, Denmark
         \and
         I. Physikalisches Institut, Universit\"at zu K\"oln, Z\"ulpicher Str. 77, 50937 K\"oln, Germany
         \and 
          Leiden Observatory, Leiden University, PO Box 9513, 2300 RA Leiden, The Netherlands
          \and
          Max-Planck Institut f\"ur Extraterrestrische Physik (MPE), Giessenbachstr. 1, 85748 Garching, Germany
             }

   \date{Received xxx; accepted xxx}

 
  \abstract
  {Nitrogen oxides are thought to play a significant role as a nitrogen reservoir and to potentially participate in the formation of more complex species. Until now, only NO, N$_2$O and HNO have been detected in the interstellar medium. We report the first interstellar detection of nitrous acid (HONO). Twelve lines were identified towards component B of the low-mass protostellar binary IRAS~16293--2422 with the Atacama Large Millimeter/submillimeter Array, at the position where NO and N$_2$O have previously been seen. A local thermodynamic equilibrium model was used to derive the column density ($\sim$\,9\,$\times$\,10$^{14}$ cm$^{-2}$ in a 0$\farcs$5 beam) and excitation temperature ($\sim$\,100\,K) of this molecule. HNO, NO$_2$, NO$^+$, and HNO$_3$ were also searched for in the data, but not detected. 
  We simulated the HONO formation using an updated version of the chemical code Nautilus and compared the results with the observations. The chemical model is able to reproduce satisfactorily the HONO, N$_2$O, and NO$_2$ abundances, but not the NO, HNO, and NH$_2$OH abundances. This could be due to some thermal desorption mechanisms being destructive and therefore limiting the amount of HNO and NH$_2$OH present in the gas phase. Other options are UV photodestruction of these species in ices or missing reactions potentially relevant at protostellar temperatures.
}
 

   \maketitle
%

\section{Introduction}

The interstellar medium is characterised by a rich and varied chemistry with closely connected groups of species found to be prominent in regions with differing physics. An example is the group of nitrogen oxides, i.e., molecules containing nitrogen-oxygen-hydrogen bonds. Secure interstellar detections have been made for three molecules: nitric oxide \citep[NO, e.g.,][]{liszt1978,mcgonagle1990,ziurys1991,caux2011,codella2018,ligterink2018}, nitrosyl hydride \citep[HNO,][]{snyder1993} and nitrous oxide \citep[N$_{2}$O,][]{ziurys1994,ligterink2018}. 
These species, in particular NO, are thought to be critical for the overall nitrogen chemistry of the ISM as they may lock up significant amounts of atomic nitrogen, and are often only second in abundance to molecular nitrogen \citep[e.g.,][]{herbstleung1986,nejad1990,pineaudesforets1990,visser2011}. Nitrogen oxides can be at the basis of greater chemical complexity, as demonstrated, for example, with the solid-state hydrogenation of NO into hydroxylamine \citep[NH$_{2}$OH,][]{Congiu2012,fedoseev2012,fedoseev2016} or energetic processing of N$_{2}$O ice
\citep{debarros2017}. 

Despite the relevance of nitrogen oxides as a nitrogen reservoir and as precursors of complex molecules, a number of important members of this group have not yet been detected in the ISM. Examples are nitrogen dioxide (NO$_{2}$), nitrous acid (HONO), and nitric acid (HNO$_{3}$), which on Earth play a role in atmospheric pollution \citep[e.g.,][]{possanzini1988}. 
In particular, the photodissociation of HONO results in abundant formation of OH radicals, which in turn engage in various oxidation reactions and the formation of ground-level ozone
\citep[O$_{3}$,][]{ren2003,lee2013,gligorovski2016,zhang2016}. Due to its relevance in atmospheric chemistry, the formation, destruction and characteristics of HONO have been well studied \citep[e.g.,][]{coxderwent1976,jenkin1988,joshi2012}. 

In this work, HONO and other nitrogen oxides were searched for towards the low-mass protostar IRAS~16293--2422 (hereafter IRAS16293), located at a distance of $\sim$140 pc in the $\rho$ Ophiuchus cloud complex \citep{Dzib2018}. This Class 0 object is known for its chemical complexity and is considered an astrochemical reference among solar-type protostars \citep[e.g.,][]{vandishoeck1995,cazaux2003,caux2011,jorgensen2016}. A large number of species have first been detected towards a low-mass source in this object. These detections include the small species NO and N$_{2}$O \citep{caux2011,ligterink2018}, the simplest ``sugar'' glycolaldehyde \citep[HOCH$_{2}$CHO,][]{jorgensen2012,jorgensen2016}, the peptide-like molecules formamide (NH$_{2}$CHO) and methyl isocyanate \citep[CH$_{3}$NCO,][]{kahane2013,coutens2016,ligterink2017,martin2017}, cyanamide \citep[NH$_2$CN,][]{coutens2018}, methyl isocyanide \citep[CH$_{3}$NC,][]{calcutt2018}, and the isomers of acetone (CH$_3$COCH$_3$) and propanal (C$_2$H$_5$CHO) as well as ethylene oxide \citep[c-C$_2$H$_4$O,][]{lykke2017}. Recently, the first interstellar detection of the organohalogen CH$_{3}$Cl was also reported towards this source \citep{fayolle2017}. 

In this letter, we present the first interstellar detection of HONO. Further constraints on the nitrogen oxide chemistry towards IRAS16293 are given, and a first attempt is made at modelling the HONO formation network.


\section{Observations and analysis}
\label{sect_obs}

Data from the Protostellar Interferometric Line Survey (PILS) of the low-mass protobinary IRAS16293 were used to search for nitrogen oxides. This survey, taken with the Atacama Large Millimeter/submillimeter Array (ALMA), is fully described in \citet{jorgensen2016}. A short overview is given here. The survey covers part of Band 7 in the spectral range 329.147--362.896 GHz, at a spectral resolution of 0.2~km~s$^{-1}$ and with a sensitivity of 6--10 mJy beam$^{-1}$ channel$^{-1}$ (i.e., 4--5 mJy beam$^{-1}$ km~s$^{-1}$). 
A circular restoring beam of 0$\farcs$5 was used to produce the final dataset. IRAS16293 is a binary. HONO is identified towards source B, but not towards source A. Source B is analysed at a position offset by one beam with respect to the continuum peak position in the south-west direction ($\alpha_{\rm J2000}$=16$^{\rm h}$32$^{\rm m}$22$\fs$58, $\delta_{\rm J2000}$=-24$\degr$28$\arcmin$32.8$\arcsec$). The very narrow line widths (1 km~s$^{-1}$) at this position limit line blending and facilitate easier identification of molecules \citep[e.g.,][]{lykke2017}. 

To analyse the spectra and identify the HONO lines, the CASSIS line analysis software\footnote{CASSIS has been developed by IRAP-UPS/CNRS (\url{http://cassis.irap.omp.eu/})}, as well as the Jet Propulsion Laboratory (JPL\footnote{\url{http://spec.jpl.nasa.gov}}) spectroscopic database \citep{Pickett1998} and the Cologne Database for Molecular Spectroscopy \citep[CDMS\footnote{\url{http://www.astro.uni-koeln.de/cdms}};][]{muller2001,muller2005} were used. The spectroscopy of HONO available in the JPL database was studied by \citet{guilmot1993a, guilmot1993b} and \citet{dehayem2005}. HONO has two different conformers, trans and cis. The JPL entry assumes that the isomers are in thermal equilibrium. The trans/cis energy difference (130.2 cm$^{-1}$) is from \citet{Varma1976}. 
Since the spectra of IRAS16293 are very line-rich, a careful check was performed to exclude blended or partially blended lines. To achieve this, we compared all lines tentatively identified as HONO with a template containing the lines of the molecules previously detected in this source (see Appendix \ref{template}). Similar to previous PILS studies (e.g., \citealt{ligterink2018}), the observed spectra were fitted with a synthetic spectrum, assuming Local Thermodynamic Equilibrium (LTE) conditions, using a source size of 0$\farcs$5 and a $V_{\rm LSR}$ velocity of 2.5~km~s$^{-1}$. As the line emission is coupled with dust emission in IRAS16293, a correction to the background temperature ($T_{\rm BG}$ = 21~K) was applied \citep[see also][]{calcutt2018,ligterink2018}. A $\chi^{2}$ minimisation routine was employed to find the best-fit model to the observed data and derive the column density ($N$) and excitation temperature ($T_{\rm ex}$, see also \citealt{lykke2017,calcutt2018,ligterink2018}). 
The grid covers excitation temperatures between 50 and 300 K with steps of 25 K. After a first estimate of the column density, the grid was refined between 5\,$\times$\,10$^{14}$ and 3\,$\times$\,10$^{15}$ cm$^{-2}$ with a step of 1\,$\times$\,10$^{14}$ cm$^{-2}$. To avoid any bias in the determination of the best-fit model with the $\chi^{2}$ calculation, we included some undetected transitions (333925.02, 348264.91, and
358979.13 MHz) that are predicted to be above the noise limit for certain models in the grid.


\section{Observational results}
\label{sect_results}

\begin{figure*}[!ht] 
\begin{center} 
\includegraphics[width=\hsize]{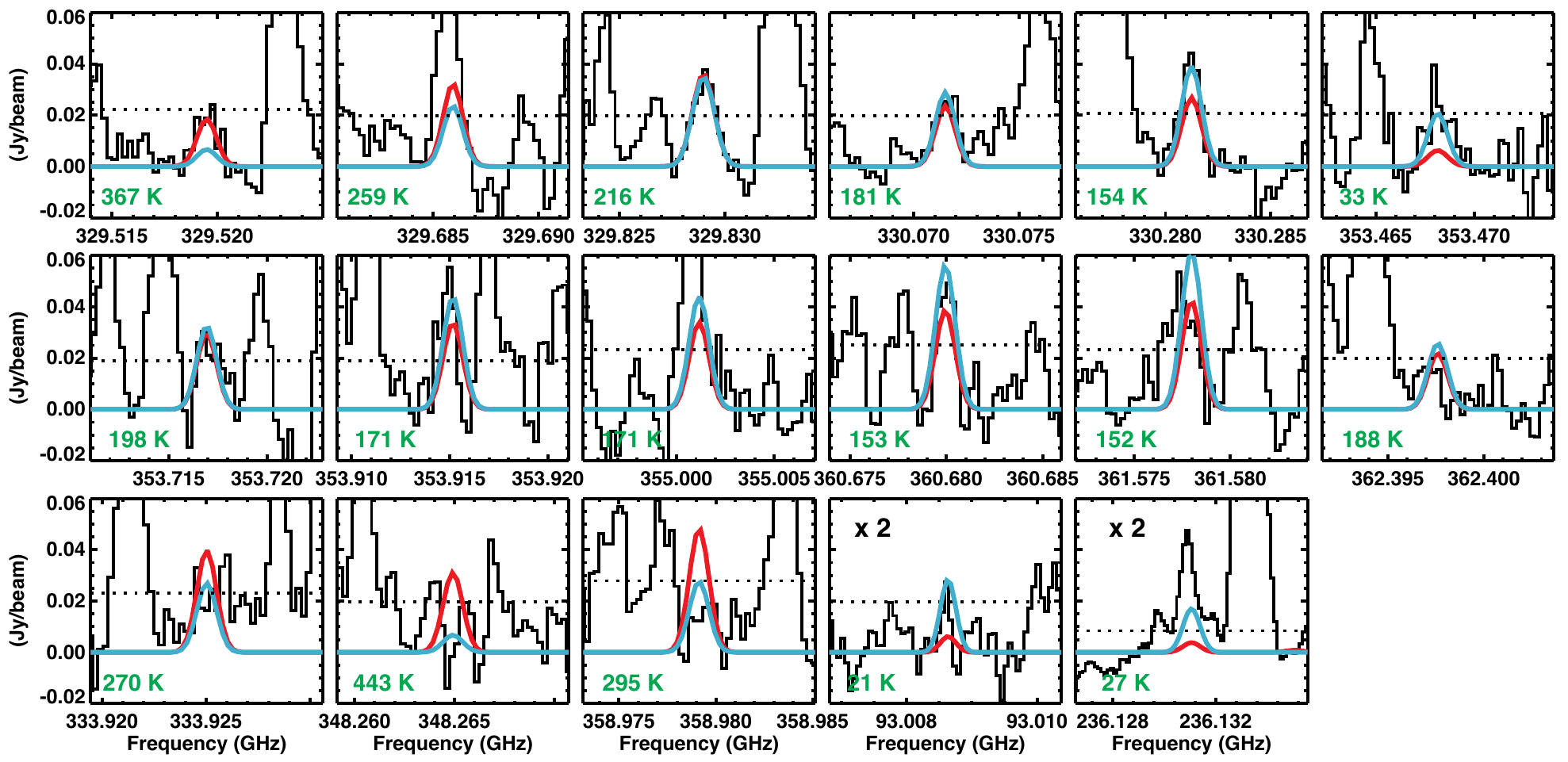}
\caption{Lines of HONO observed towards the protostar IRAS16293 B (in black). The first 12 lines are the identified lines of HONO in the ALMA-PILS band 7 survey. On the last row, the first 3 lines correspond to the undetected transitions that are used to constrain the best-fit model and the last two lines are the ones identified in other ALMA data (see Section \ref{sect_results} for more details). The 3$\sigma$ limit is indicated by a dotted line. The best-fit model with $T_{\rm ex}$ = 100 K is in blue, while the model in red corresponds to a higher $T_{\rm ex}$ of 300 K. The spectrum at 93 GHz is extracted at the continuum peak position, given the lower spatial resolution of the data. The column density was multiplied by a factor 2 to take this difference into account \citep{jorgensen2016}. The upper energy level is indicated in green in the lower left corner of each panel.}
\label{fig_obs} 
\end{center} 
\end{figure*} 
 
  \begin{figure}[!ht] 
 \begin{center} 
 \includegraphics[width=\hsize]{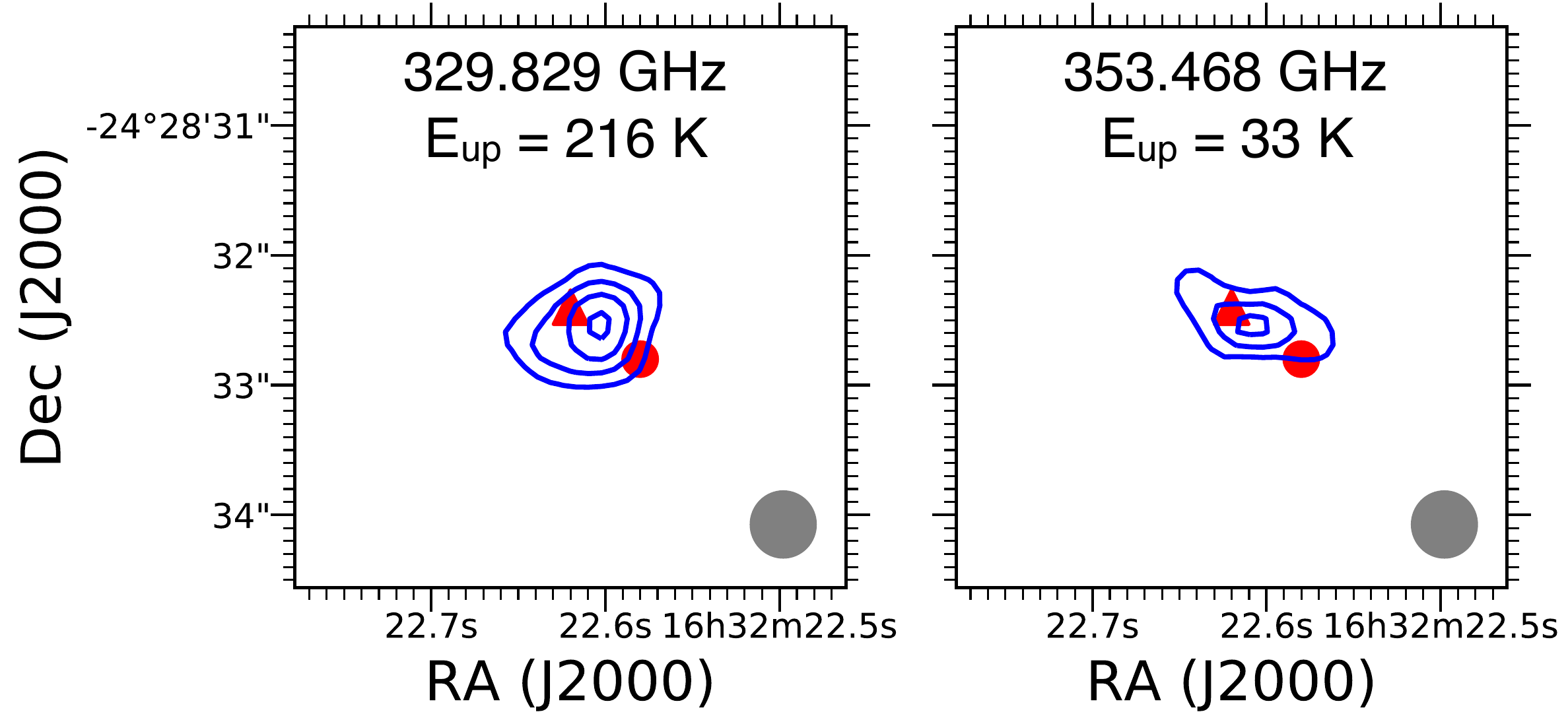}
 \caption{Integrated intensity maps of two transitions of HONO towards IRAS16293 B. The position of the continuum peak is indicated with a red triangle, while the position analysed for IRAS16293 B (full-beam offset) is indicated with a red circle. The beam size is indicated in grey in the bottom right corner. Left: contours are 5, 10, 15, and 20$\sigma$. Right: contour levels are 3, 6, and 9$\sigma$. }
 \label{fig_map}
  \end{center} 
 \end{figure}  
 
In total, we found 12 lines that could be identified as (trans-) HONO, that are not blended with any known species (see Figure \ref{fig_obs} and Table \ref{tab.lines}). The intensities of nine out of these lines are higher or equal to 5$\sigma$. Two lines are 3 or 4$\sigma$ detections and one is a marginal (2$\sigma$) detection. The best-fit model is obtained for an excitation temperature of 100 K and a column density of 9\,$\times$\,10$^{14}$ cm$^{-2}$. The column density is not very sensitive to the excitation temperature. For a fixed excitation temperature of 300 K (which is derived for several complex organic molecules, see \citealt{jorgensen2018}), the best-fit column density is 1.4\,$\times$\,10$^{15}$ cm$^{-2}$, i.e. only 50\% larger. Nevertheless, the model at 300 K overproduces some undetected lines at 333925.02, 348264.91, and 358979.13 MHz (Table \ref{tab.lines_und}) and does not properly reproduce the line at 353468.14 MHz, although it better reproduces the line at 329519.48 MHz than the model at 100 K (see Figure \ref{fig_obs}). The best-fit excitation temperature of 100 K is consistent with the excitation temperature obtained for the other nitrogen oxides, especially NO \citep{ligterink2018}. Three lines (329519.48, 329685.92, and 355001.15 GHz) have their fluxes underproduced by the best-fit model and could be blended with unknown species, though the first one is only detected at 3$\sigma$. Alternatively, it could be that for molecules with low-frequency vibrational modes like HONO, the excitation does not need to be in LTE, but there could be infrared-pumping for selected lines.

Lines of HONO were also searched towards other high sensitivity ALMA observations of the low-mass protostar IRAS16293. One line is present at 93008.6 MHz in the lower spatial resolution data of the ALMA-PILS observations carried out in band 3 \citep[][see Figure \ref{fig_obs}]{jorgensen2016}. None are present in the band 6 data. 
According to our calculations, one HONO transition at 236131.076 MHz should also be observed in the ALMA data presented in \citet{taquet2018} with an intensity of 7 mJy for a similar spatial resolution of 0$\farcs$5. An unidentified line is present at the same frequency, but its intensity is a factor 3 higher, which could mean that the observed line is blended with another species (see Figure \ref{fig_obs}). 

 Maps of HONO (see Figure \ref{fig_map}) show that the emission is quite compact around IRAS16293 B, similarly to the majority of the molecules detected in this source, especially the complex organic molecules (see e.g., \citealt{coutens2016,coutens2018,lykke2017}) and NO \citep{ligterink2018}.

Four other nitrogen-oxides, nitrosyl hydride (HNO), nitrosyl cation (NO$^{+}$), nitrogen dioxide (NO$_{2}$), and nitric acid (HNO$_
3$) were searched for, but not identified (see Appendix \ref{ap.upplim} for details). Table \ref{tab:coldens} gives an overview of the derived column densities of the detected and unidentified species (upper limits) towards IRAS16293 B, and includes results on NO, N$_{2}$O, and NH$_{2}$OH from \citet{ligterink2018}.

\begin{table}[!t]
\footnotesize
\caption{Column densities at the one beam offset position of IRAS16293 B from the ALMA-PILS data. \label{tab:coldens}}
\center
\begin{tabular}{llccc}
\hline
\hline
Molecule & Formula & $N_\mathrm{tot}^{\dagger}$ (cm$^{-2}$) & $T_\mathrm{ex}$ (K) \\
\hline
Nitrous acid & HONO & (9$\pm$5)\,$\times$\,10$^{14}$ & 100 \\
Nitric oxide$^{\ddagger}$ & NO & \phantom{$\leq$}(2.0$\pm$0.5)\,$\times$\,10$^{16}$ & 40--150  \\
Nitrous oxide$^{\ddagger}$ & N$_2$O & $\geq$ 4.0\,$\times$\,10$^{16}$ & 25--350 \\ 
Hydroxylamine$^{\ddagger}$ & NH$_2$OH & $\leq$ 4\,$\times$\,10$^{14}$ & [100] \\
Nitrosyl hydride & HNO & $\leq$ 3\,$\times$\,10$^{14}$ & [100] \\
Nitrogen dioxide & NO$_{2}$ & $\leq$ 2\,$\times$\,10$^{16}$  & [100] \\
Nitrosyl cation & NO$^{+}$ & $\leq$ 2\,$\times$\,10$^{14}$ & [100] \\
Nitric acid & HNO$_3$ & $\leq$ 5\,$\times$\,10$^{14}$ & [100] \\
\hline
\end{tabular}
\\
\tablefoot{All models assume LTE, a $FWHM$ of 1\,km\,s$^{-1}$, a peak velocity $V_{\rm peak}$ of 2.5$\pm$0.2 km\,s$^{-1}$ and a source size of 0\farcs5. $^{\dagger}$The uncertainties are 3$\sigma$. Upper limits are also 3$\sigma$ and determined for an assumed $T_{\rm ex}$ = 100~K, indicated with brackets in the table. $^{\ddagger}$Results from \citet{ligterink2018}. } 
\end{table}

\begin{table*}[!ht]
\footnotesize
\caption{Abundances of HONO derived in IRAS16293 B and with the chemical model \label{tab:ab}}
\center
\begin{tabular}{lccccccc}
\hline
\hline
 & [HONO]/[NO] & [HONO]/[N$_{2}$O] & [HONO]/[HNO] & [HONO]/[NO$_{2}$] & [HONO]/[NH$_{2}$OH] & [HONO]/[CH$_{3}$OH] \\
\hline
IRAS16293 B & 4.5\,$\times$\,10$^{-2}$ & $\leq$2.3\,$\times$\,10$^{-2,\dagger}$ & $\geq$3 & $\geq$4.5\,$\times$\,10$^{-2}$ & $\geq$2.3 & 9\,$\times$\,10$^{-5}$\\
Chemical model & 71 & 2\,$\times$\,10$^{-2}$ & 1\,$\times$\,10$^{-2}$ & 533 & 1.7\,$\times$\,10$^{-3}$ & $1.9\times 10^{-4}$ \\
\hline
\end{tabular}
\tablefoot{$^{\dagger}$Upper limit due to a lower limit on N$_{2}$O.}
\end{table*}


\section{Chemical modeling of HONO}
\label{sect_model}

To describe HONO and other N$_{x}$O$_{y}$H$_{z}$ species, we have updated the gas and grain chemical network used in \citet[][and references therein]{Loison2019} introducing various species such as HONO, s-HONO, s-HNO$_2$, s-NH$_2$O, s-HNOH, NH$_2$OH, s-NH$_2$OH, s-NO$_3$, s-HNO$_3$, s-H$_3$NO$_2$ (s- indicates species on grains). The reactions involving HONO are summarised in Table C.1 (we do not present the full network in this letter). The chemistry of HONO is well described for Earth atmosphere chemistry where it is produced through the barrierless three body OH + NO + M reaction  \citep{Forster1995,Atkinson2004}. However, this reaction is inefficient at the low densities of interstellar clouds and the radiative rate constant is negligible due to the small size of the system. All other known gas phase reactions producing HONO have negligible rates in the interstellar medium. In our model, HONO is therefore produced on grains through the s-O + s-HNO, s-H + s-NO$_2$, and s-OH + s-NO surface reactions, all of which are barrierless in the gas phase \citep{Inomata1999,Du2004,Michael1979,Nguyen1998,Su2002,Forster1995,Atkinson2004}. 

This network was then used with the Nautilus gas-grain model \citep{2016MNRAS.459.3756R}, which computes the gas and grain chemistry. The chemical modeling was done in two steps as in similar previous studies of IRAS16293 \citep[see for instance][]{Andron2018}: a cold core phase (a gas and dust temperature of 10~K, an atomic H density of $10^4$ cm$^{-3}$, a visual extinction ($A_{\rm V}$) of 15, and a cosmic-ray ionisation rate of $1.3\times 10^{-17}$~s$^{-1}$) during $10^6$~yrs followed by a collapse phase. 
For the collapse, we have used the physical structure derived from a 1D radiative hydrodynamical model \citep[see][]{2008ApJ...674..984A} for parcels of material collapsing towards the central star. For these simulations, we have used these parcels arriving at 62.4 au at the end of the simulations \citep[see Fig. 5 of][]{2008ApJ...674..984A}. The resulting abundance ratios at this radius at the end of the simulation are presented in Table \ref{tab:ab}. 

In our model, HONO is essentially formed during the cold core phase. The final HONO/CH$_3$OH ratio predicted by the model is close to the observed value within a factor of 2. The model ratios HONO/N$_2$O and HONO/NO$_2$ are also in agreement with the observed upper and lower limits, respectively. However, our model produces too little NO and too much HNO at high temperatures in the gas phase, resulting in a HONO/NO ratio much larger and a HONO/HNO ratio much smaller than the respective observed values. In our model, most of the NO reacts on grains with other radicals such as s-NH, when the temperature increases and NO becomes mobile. An explanation for the large NO/HNO ratio observed in IRAS16923 B could be that thermal desorption of s-HNO mainly results in its destruction to NO, due to the weak H-NO bond of HNO (2.02 eV, \citealt{Dixon1996}), as the formation of s-HNO is very likely due to the absence of a barrier for the s-H + s-NO reaction \citep{Tsang1991, Nguyen2004,Washida1978,Glarborg1998}. It should be noted that our model (as well as other published ones) overproduces the abundance of NH$_2$OH, a molecule so far not detected in the interstellar medium \citep{Pulliam2012,McGuire2015,ligterink2018}. It has been suggested that NH$_2$OH cannot desorb without destruction by \citet{Jonusas2016}, although this is in contradiction with the laboratory experiments of \citet{Congiu2012}.  Despite their differences, both experimental studies used very similar Temperature Programmed Desorption (TPD) setups and new experiments are therefore clearly needed to address these discrepancies. Other processes such as UV-photodestruction of these species in ices could also explain the discrepancy between the model and the observations \citep{fedoseev2016}. In addition, the chemical network on grains and in the gas phase may not be fully relevant at protostellar temperatures. Some reactions with barriers, absent from the current network, may be significant.

\section{Conclusions}
\label{sect_conclusion}

We report the first detection of nitrous acid (HONO) in the interstellar medium. This molecule, which is known to play a major role in the atmosphere of our planet, was found with ALMA towards the well-studied solar-type protostar IRAS16293~B. This discovery complements the recent detection of N$_2$O in the same source \citep{ligterink2018} and expands our knowledge of the chemical network of nitrogen oxides. Our updated model allows the abundances of HONO, N$_2$O, and NO$_2$ to be reproduced satisfactorily, but not the ones of NO, HNO, and NH$_2$OH. One reason could be that HNO and NH$_2$OH are destroyed upon thermal desorption, something which deserves to be experimentally studied in detail. Other explanations could be that they are destroyed by UV photons in ices or that some grain surface or gas-phase reactions, potentially relevant at protostellar temperatures, are missing from the network.

\begin{acknowledgements}

This paper makes use of the ALMA data ADS/JAO.ALMA\#2013.1.00278.S. ALMA is a partnership of ESO (representing its member states), NSF (USA) and NINS (Japan), together with NRC (Canada) and NSC and ASIAA (Taiwan), in cooperation with the Republic of Chile. The Joint ALMA Observatory is operated by ESO, AUI/NRAO and NAOJ. A.C. postdoctoral grant is funded by the ERC Starting Grant 3DICE (grant agreement 336474). V.W. and J.-C.L. acknowledge the CNRS program Physique et Chimie du Milieu Interstellaire (PCMI) co-funded by the Centre National d’Etudes Spatiales (CNES). M.N.D. acknowledges the financial support of the SNSF Ambizione grant 180079, the Center for Space and Habitability (CSH) Fellowship and the IAU Gruber Foundation Fellowship. 
J.K.J. acknowledges support from ERC Consolidator Grant ``S4F'' (grant agreement 646908). Research at the Centre for Star and Planet Formation is funded by the Danish National Research Foundation.
\end{acknowledgements}

\bibliographystyle{aa} 
\bibliography{Biblio} 



\appendix

\section{Lines of HONO}
\label{template}

\begin{figure*}[!p] 
 \begin{center} 
 \includegraphics[width=\hsize]{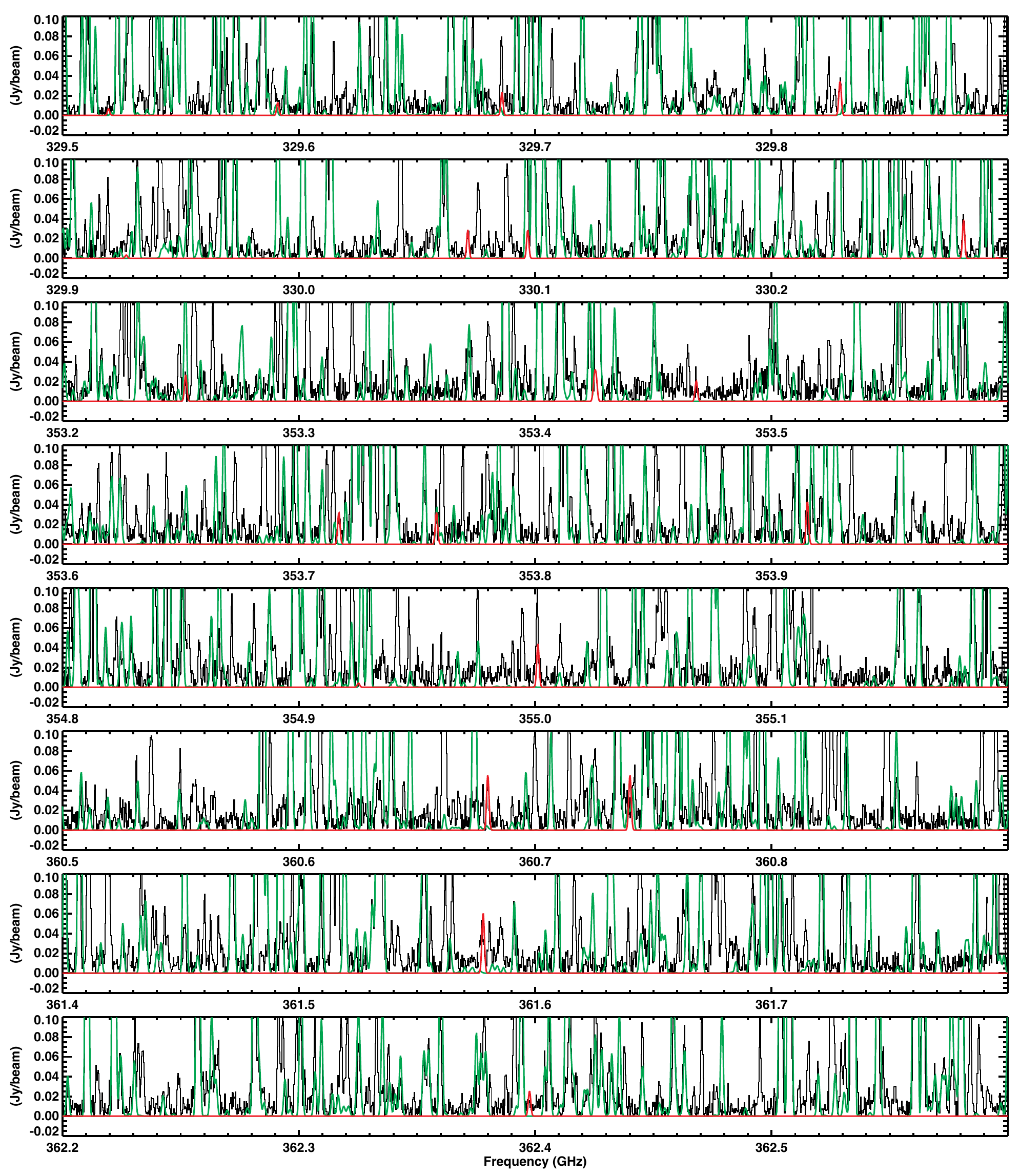}
 \caption{Lines of HONO observed towards the protostar IRAS16293 B over a larger spectral range. The HONO best-fit model is shown in red, while the template model used to check the potential blending of the lines is overlaid in green.}
 \label{fig_template}
  \end{center} 
 \end{figure*}

The detected HONO transitions that are not found to be blended with known species are listed in Table \ref{tab.lines}.  To check the potential blending of the HONO lines with other species, we defined a template based on the molecules previously identified in the ALMA-PILS survey. This template includes the following species (ranked by mass): CCH, HCN, HNC, H$^{13}$CN, HC$^{15}$N, DNC, CO, $^{13}$CO, C$^{17}$O, H$^{13}$C$^{15}$N, CH$_2$NH, NO, C$^{18}$O, DCO$^+$, H$_2$CO, HDCO, H$_2^{13}$CO, H$_2$C$^{17}$O, D$_2$CO, H$_2$C$^{18}$O, CH$_3$OH, CH$_2$DOH, CH$_3$OD,  $^{13}$CH$_3$OH, D$_2^{13}$CO, H$_2$S, CH$_3^{18}$OH, HDS, HD$^{34}$S, c-C$_3$H$_2$, CH$_3$CCH, CH$_3$CN, CH$_3$NC, NH$_2$CN, H$_2$CCO, $^{13}$CH$_3$CN, CH$_3^{13}$CN, CH$_3$C$^{15}$N, CH$_2$DCN, H$_2$C$^{13}$CO, H$_2^{13}$CCO, HDCCO, HNCO, CHD$_2$CN, NH$_2^{13}$CN, NHDCN, CH$_3$CHO, N$_2$O, DNCO, HN$^{13}$CO, CS, c-C$_2$H$_4$O, SiO, CH$_3$CDO, $^{13}$CH$_3$CHO, CH$_3^{13}$CHO, C$^{33}$S, NH$_2$CHO, C$^{34}$S, t-HCOOH, H$_2$CS, NH$_2^{13}$CHO, cis-NHDCHO, trans-NHDCHO, NH$_2$CDO, CH$_3$OCH$_3$, C$_2$H$_5$OH, DCOOH, HCOOD, t-H$^{13}$COOH, HDCS, a-CH$_3^{13}$CH$_2$OH, a-$^{13}$CH$_3$CH$_2$OH, a-CH$_3$CH$_2$OD, a-CH$_3$CHDOH, a-a-CH$_2$DCH$_2$OH, a-s-CH$_2$DCH$_2$OH, $^{13}$CH$_3$OCH$_3$, a-CH$_2$DOCH$_3$, sym-CH$_2$DOCH$_3$, SO, C$^{36}$S, CH$_3$SH, CH$_3^{35}$Cl, HC$_3$N, CH$_3^{37}$Cl,  C$_2$H$_3$CN, C$_2$H$_5$CN, CH$_3$NCO, CH$_3$COCH$_3$, C$_2$H$_5$CHO, CH$_3$OCHO, CH$_2$(OH)CHO, OCS, CH$_3$COOH, O$^{13}$CS, OC$^{33}$S, CH$_2$(OH)$^{13}$CHO, $^{13}$CH$_2$(OH)CHO, CH$_3$O$^{13}$CHO, CH$_2$(OD)CHO, CHD(OH)CHO, CH$_2$(OH)CDO, CH$_3$OCDO, CH$_2$DOCHO, CHD$_2$OCHO,  aGg'-(CH$_2$OH)$_2$, gGg'-(CH$_2$OH)$_2$, OC$^{34}$S, $^{18}$OCS, CHD$_2$OCHO, SO$_2$, and $^{34}$SO$_2$. 
Figure \ref{fig_template} presents the HONO lines over a larger spectral range with the overlaid template model in green and the HONO model in red. 

\begin{table*}[!ht]
\caption{List of the detected and unblended HONO transitions. \label{tab.lines}}
\begin{tabular}{rcccccc}
\hline
\hline
Molecule & Transition & Frequency & $E_{\rm up}$ & $A_{\rm ij}$ & $g_{\rm up}$ \\
 & (MHz) & (K) & (s$^{-1}$) & \\
 \hline
trans-HONO & 14 8 6 -- 13 8 5 & 329519.5 & 367.2 & 2.57\,$\times$\,10$^{-4}$ & 29 \\
trans-HONO & 14 8 7 -- 13 8 6 & 329519.5 & 367.2 & 2.57\,$\times$\,10$^{-4}$ & 29 \\
trans-HONO & 14 6 8 -- 13 6 7 & 329685.9 & 258.6 & 3.12\,$\times$\,10$^{-4}$ & 29 \\
trans-HONO & 14 6 9 -- 13 6 8 & 329685.9 & 258.6 & 3.12\,$\times$\,10$^{-4}$ & 29 \\
trans-HONO & 14 5 10 -- 13 5 9 & 329828.8 & 215.9 & 3.34\,$\times$\,10$^{-4}$ & 29 \\
trans-HONO & 14 5 9 -- 13 5 8 & 329829.2 & 215.9 & 3.34\,$\times$\,10$^{-4}$ & 29 \\
trans-HONO & 14 4 11 -- 13 4 10 & 330071.5 & 181.0 & 3.52\,$\times$\,10$^{-4}$ & 29 \\
trans-HONO & 14 3 12 -- 13 3 11 & 330281.2 & 153.8 & 3.67\,$\times$\,10$^{-4}$ & 29 \\
trans-HONO & 5 2 4 -- 4 1 3 & 353468.1 & 32.5 & 1.50\,$\times$\,10$^{-4}$ & 11 \\
trans-HONO & 15 4 12 -- 14 4 11 & 353716.9 & 198.0 & 4.40\,$\times$\,10$^{-4}$ & 31 \\
trans-HONO & 15 3 13 -- 14 3 12 & 353915.2 & 170.8 & 4.55\,$\times$\,10$^{-4}$ & 31 \\
trans-HONO & 15 3 12 -- 14 3 11 & 355001.2 & 171.0 & 4.59\,$\times$\,10$^{-4}$ & 31 \\
trans-HONO & 15 2 13 -- 14 2 12 & 360679.9 & 152.9 & 4.93\,$\times$\,10$^{-4}$ & 31 \\
trans-HONO & 16 1 16 -- 15 1 15 & 361578.0 & 151.9 & 5.04\,$\times$\,10$^{-4}$ & 33 \\
trans-HONO & 17 2 16 -- 17 1 17 & 362397.6 & 187.8 & 2.77\,$\times$\,10$^{-4}$ & 35 \\
\hline
\end{tabular}        
\tablefoot{Quantum numbers are given as $J'$\,$K_{a}'$\,$K_{c}'$ -- $J''$\,$K_{a}''$\,$K_{c}''$.}
\end{table*}

\begin{table*}[!ht]
\caption{List of the undetected HONO transitions used in the $\chi^{2}$ calculation. \label{tab.lines_und}}
\begin{tabular}{r c c c c c c}
\hline
\hline
Species & Transition & Frequency & $E_{\rm up}$ & $A_{\rm ij}$ & $g_{\rm up}$ \\
  & (MHz) & (K) & (s$^{-1}$) & \\
 \hline
trans-HONO & 21 1 20 -- 21 0 21 & 333925.0 & 270.3 & 5.42\,$\times$\,10$^{-4}$ & 43 \\
cis-HONO & 20 1 19 -- 20 0 20 & 348264.9 & 443.2 & 7.87\,$\times$\,10$^{-4}$ & 41 \\
trans-HONO & 22 1 21 -- 22 0 22 & 358979.1 & 295.3 & 6.82\,$\times$\,10$^{-4}$ & 45 \\
\hline
\end{tabular}  
\tablefoot{Quantum numbers are given as $J'$\,$K_{a}'$\,$K_{c}'$ -- $J''$\,$K_{a}''$\,$K_{c}''$.}
\end{table*}

\section{Upper limit determination of HNO, NO$^{+}$, NO$_{2}$, and HNO$_3$}
\label{ap.upplim}

The molecules HNO, NO$^{+}$, NO$_{2}$, and HNO$_3$ were searched for in the PILS data without success. 3$\sigma$ upper limit column densities were determined for $T_{\rm ex}$ = 100~K. When possible, line-free regions of the observed data, where transitions are expected, were used. 
For HNO, the 4$_{1,3}$ -- 3$_{1,2}$ transition at 332106.6 MHz, which is close to the transition of an unidentified species, was used. 
For NO$^{+}$, only the 3--2 transition at 357.564 GHz is covered in the PILS range. It is found to be blended with a line of ethylene glycol. We determined a conservative upper limit based on the total flux  of this line. For NO$_{2}$, the undetected transitions at 348062.5 and 348820.7 MHz were used. For HNO$_3$, many undetected lines are covered in the spectral range of the PILS survey. The upper limit was derived based on the brightest lines predicted in line-free regions. 
At $T_{\rm ex}$ = 100~K, this yields 3$\sigma$ upper limit column densities of 3\,$\times$\,10$^{14}$, 2\,$\times$\,10$^{14}$, 2\,$\times$\,10$^{16}$, and 5\,$\times$\,10$^{14}$ cm$^{-2}$ for HNO, NO$^{+}$, NO$_{2}$, and HNO$_3$, respectively.

\section{Chemical reactions for HONO}

The reactions involving HONO are listed in Table C.1.

\onecolumn

\begin{sidewaystable}[]
\footnotesize
\caption{Summary of the reactions involving HONO \label{tab:reactions}}
\begin{tabular}{|r|ll|l|l|l|l|l|l|p{0.4\linewidth}|}
\hline
 & Reaction &  & \begin{tabular}[c]{@{}l@{}}$\Delta$E \\ kJ/mol\end{tabular} & $\alpha$ & $\beta$ & $\gamma$ & F$_{0}$ & g & Reference \\ \hline
1. & \multicolumn{1}{l}{\begin{tabular}[c]{@{}r@{}}H$^{+}$ + HONO\\ -\end{tabular}} & \begin{tabular}[c]{@{}l@{}}$\rightarrow$ HONO$^{+}$ + H\\ $\rightarrow$ H$_{2}$O + NO$^{+}$\end{tabular} & \begin{tabular}[c]{@{}l@{}}-235\\ -693\end{tabular} & \begin{tabular}[c]{@{}l@{}}0\\ 1.0\end{tabular} & \begin{tabular}[c]{@{}l@{}}\\ 2.0$\times$10$^{-9}$\end{tabular} & \begin{tabular}[c]{@{}l@{}}\\ 5.27\end{tabular} & \begin{tabular}[c]{@{}l@{}}\\ 3\end{tabular} & \begin{tabular}[c]{@{}l@{}} \\ 0\end{tabular} & Ionpol1, capture rate theory. We avoid introducing HONO$^{+}$. \\ \hline
2. & \begin{tabular}[c]{@{}l@{}}He$^{+}$ + HONO\\ \\ \end{tabular} & \begin{tabular}[c]{@{}l@{}}$\rightarrow$ HONO$^{+}$ + He\\ $\rightarrow$ HO + NO$^{+}$ +  He\end{tabular} &  & \begin{tabular}[c]{@{}l@{}}0\\ 1.0\end{tabular} & \begin{tabular}[c]{@{}l@{}}\\ 1.0$\times$10$^{-9}$\end{tabular} & \begin{tabular}[c]{@{}l@{}}\\ 5.27\end{tabular} & \begin{tabular}[c]{@{}l@{}}\\ 3\end{tabular} & \begin{tabular}[c]{@{}l@{}}\\ 0\end{tabular} & Ionpol1, capture rate theory. We avoid introducing HONO$^{+}$. \\ \hline
3. & C + HONO & $\rightarrow$ CO + NO + H & -459 & 3.0$\times$10$^{-10}$ & 0 & 0 & 2 & 0 & Capture rate theory, approximate branching ratio. \\ \hline
4. & \begin{tabular}[c]{@{}l@{}}C$^{+}$ + HONO\\ \\ \end{tabular} & \begin{tabular}[c]{@{}l@{}}$\rightarrow$ CO + NO$^{+}$ + H\\ $\rightarrow$ HCO$^{+}$ + NO\end{tabular} & \begin{tabular}[c]{@{}l@{}}-635\\ -817\end{tabular} & \begin{tabular}[c]{@{}l@{}}2.0$\times$10$^{-9}$\\ 0\end{tabular} & \begin{tabular}[c]{@{}l@{}}-0.4\\ -0.4\end{tabular} & \begin{tabular}[c]{@{}l@{}}0\\ 0\end{tabular} & \begin{tabular}[c]{@{}l@{}}3\\ 3\end{tabular} & \begin{tabular}[c]{@{}l@{}}0\\ 0\end{tabular} & \vspace{-0.4cm}
Capture rate theory, approximate branching ratio. (HCO$^{+}$ + NO is likely a non-negligible exit channel). \\ \hline
5. & OH + HONO & $\rightarrow$ NO$_{2}$ + H$_{2}$O & -21 & 7.0$\times$10$^{-12}$ & -0.6 & 0 & 1.6 & 10 & This reaction has been studied between 278 and 373K. The results from \citet{jenkin1987} show large uncertainty due to complicated secondary reactions. The results from \citet{burkholder1992} show no barrier and a negative temperature dependency. We use an expression compatible with experimental data and leading to reasonable rate constant at 10K. \\ \hline
6. & \begin{tabular}[c]{@{}l@{}}HONO + H$^{+}_{3}$\\ \\ \\ \end{tabular} & \begin{tabular}[c]{@{}l@{}}$\rightarrow$ H$_{2}$ONO$^{+}$ + H$_{2}$\\ $\rightarrow$ HONOH$^{+}$ + H$_{2}$\\ $\rightarrow$ HONHO$^{+}$ + H$_{2}$\end{tabular} & \begin{tabular}[c]{@{}l@{}}-358\\ -232\\ -196\end{tabular} & \begin{tabular}[c]{@{}l@{}}1.0\\ 0\\ 0\end{tabular} & \begin{tabular}[c]{@{}l@{}}2.38$\times$10$^{-9}$\\ \\ \\ \end{tabular} & \begin{tabular}[c]{@{}l@{}}5.27\\ \\ \\ \end{tabular} & \begin{tabular}[c]{@{}l@{}}2\\ \\ \\ \end{tabular} & \begin{tabular}[c]{@{}l@{}}0\\ \\ \\ \end{tabular} & \vspace{-0.6cm} Ionpol1. The others isomers are likely also produced but we avoid introducing too many species without notable different chemical behavior. \\ \hline
7. & HONO + HCO$^{+}$ & $\rightarrow$ H$_{2}$ONO$^{+}$ + CO & -210 & 1.0 & 9.45$\times$10$^{-10}$ & 5.27 & 2 & 0 & Ionpol1. \\ \hline
8. & \begin{tabular}[c]{@{}l@{}}H$_{2}$ONO$^{+}$ + e$^{-}$\\ \\ \\ \\ \end{tabular} & \begin{tabular}[c]{@{}l@{}}$\rightarrow$ HONO + H\\ $\rightarrow$ H$_{2}$O + NO\\ $\rightarrow$ H + OH + NO\\ $\rightarrow$ H$_{2}$O + N + O\end{tabular} & \begin{tabular}[c]{@{}l@{}}-524\\ -825\\ -339\\ -207\end{tabular} & \begin{tabular}[c]{@{}l@{}}2.0$\times$10$^{-7}$\\ 1.0$\times$10$^{-7}$\\ 2.0$\times$10$^{-7}$\\ 1.0$\times$10$^{-7}$\end{tabular} & \begin{tabular}[c]{@{}l@{}}-0.5\\ -0.5\\ -0.5\\ -0.5\end{tabular} & \begin{tabular}[c]{@{}l@{}}0\\ 0\\ 0\\ 0\end{tabular} & \begin{tabular}[c]{@{}l@{}}3\\ 3\\ 3\\ 3\end{tabular} & \begin{tabular}[c]{@{}l@{}}0\\ 0\\ 0\\ 0\end{tabular} & \vspace{-0.8cm} Rate by comparison with similar DR \citep{fournier2013,florescu2006,Geppert2004} and branching ratios deduced roughly from similar reactions using \citet{plessis2012}. The important fact is that HONO is likely non negligible product. \\ \hline
9. & \begin{tabular}[c]{@{}l@{}}s-H + s-NO$_{2}$\\ \\ \\ \end{tabular} & \begin{tabular}[c]{@{}l@{}}$\rightarrow$ s-HONO\\ $\rightarrow$ HNO$_{2}$\\ $\rightarrow$ s-NO + s-OH\end{tabular} & \begin{tabular}[c]{@{}l@{}}-317\\ -281\\ -132\end{tabular} & \begin{tabular}[c]{@{}l@{}}0.7\\ 0.3\\ 0\end{tabular} & \begin{tabular}[c]{@{}l@{}}0\\ 0\\ \\ \end{tabular} &  &  &  & \vspace{-0.6cm} Radical-radical reaction, well known in gas phase \citep{Michael1979,Nguyen1998,Su2002}. Both approach towards N and O atoms are attractive. \\ \hline
10. & \begin{tabular}[c]{@{}l@{}}s-H + s-HONO\\ \\ \\ \\ \end{tabular} & \begin{tabular}[c]{@{}l@{}}$\rightarrow$ s-H$_{2}$NO$_{2}$\\ $\rightarrow$ s-H$_{2}$ + s-NO$_{2}$\\ $\rightarrow$ s-NO + s-H$_{2}$O\\ $\rightarrow$ s-HON + s-OH\end{tabular} & \begin{tabular}[c]{@{}l@{}}-163\\ -109\\ -301\\ +164\end{tabular} & \begin{tabular}[c]{@{}l@{}}1\\ 0\\ 0\\ 0\end{tabular} & \begin{tabular}[c]{@{}l@{}}2600\\ \\ \\ \\ \end{tabular} &  &  &  & \vspace{-0.7cm} We use the theoretical work from \citet{hsu1997}. H$_{2}$NO$_{2}$ = (HN(O)OH). Some NO + H$_{2}$O are likely to be produced but this exit channel involves a transition state close to the H + HONO entrance level. \\ \hline
11. & \begin{tabular}[c]{@{}l@{}}s-H + s-H$_{2}$NO$_{2}$\\ \\ \\ \end{tabular} & \begin{tabular}[c]{@{}l@{}}$\rightarrow$ s-H$_{3}$NO$_{2}$\\ $\rightarrow$ s-HON + s-H$_{2}$O\\ $\rightarrow$ s-HNO + s-H$_{2}$O\end{tabular} & \begin{tabular}[c]{@{}l@{}}-302\\ -159\\ -327\end{tabular} & \begin{tabular}[c]{@{}l@{}}1\\ 0\\ 0\end{tabular} & \begin{tabular}[c]{@{}l@{}}0\\ \\ \\ \end{tabular} &  &  &  & \vspace{-0.5cm} M06-2X/AVTZ calculations (this work). H$_{3}$NO$_{2}$ = HO-NH-OH. Some HNO may be produced. \\ \hline
12. & s-OH + s-NO & $\rightarrow$ s-HONO & -185 & 1 & 0 &  &  &  & The OH + NO reaction is a radical-radical barrierless reaction \citep{Forster1995}. \\ \hline
\end{tabular}
\tablefoot{The exothermicity of the reaction ($\Delta$E in kJ/mol) are calculated at M06-2X/AVTZ level using Gaussian 2009 software.
Definitions of $\alpha$, $\beta$, $\gamma$, F$_{0}$, g, Ionpol1 and Ionpol2 can been found in \citet{wakelam2010,wakelam2012}:
k = $\alpha$ $\times$ (T/300)$^{\beta}$ $\times$ exp(-$\gamma$/T) cm$^{3}$ molecule$^{-1}$ s$^{-1}$, T range is 10-300K except in some cases (noted).
Ionpol1: k = $\alpha\beta$(0.62+0.4767$\gamma$ (300/T)$^{0.5}$) cm$^{3}$ molecule$^{-1}$ s$^{-1}$,
Ionpol2: k = $\alpha\beta$(1+0.0967$\gamma$ (300/T)$^{0.5}$+($\gamma^{2}$/10.526) (300/T)) cm$^{3}$ molecule$^{-1}$ s$^{-1}$,
F$_{0}$ = exp($\Delta$k/k$_{0}$) ($\approx$1+($\Delta$k/k$_{0}$) and F(T)=F$_{0}$ exp(g $\times$ abs(1/T-1/T$_0$)).}
\end{sidewaystable}

\end{document}